\documentclass[11pt]{article}
\usepackage{graphicx}
\usepackage[margin=1.25in]{geometry}
\usepackage[usenames,dvipsnames]{color}
\usepackage{subfigure}
\newcommand{\comment}[1]{}
\usepackage{floatrow}
\usepackage{url}
\usepackage[colorlinks = true,
            linkcolor = blue,
            urlcolor  = blue,
            citecolor = blue,
            anchorcolor = blue]{hyperref}

\newcommand\pubnumber{ANL-HEP-173859\\MPP-2022-28}

\newcommand\pubdate{\today}

\def\Title#1{\begin{center} {\LARGE #1 } \end{center}}
\def\Author#1{\begin{center}{ \sc #1} \end{center}}
\def\Address#1{\begin{center}{ \it #1} \end{center}}

\newcommand\pubblock{\rightline{\begin{tabular}{l} \pubnumber\\
         \pubdate \end{tabular}}}
\newenvironment{Abstract}{\begin{quotation} \begin{center}
                       ABSTRACT
     \end{center}\bigskip  }{\end{quotation}}

\def\beq{\begin{equation}}
\def\eeq#1{\label{#1}\end{equation}}
\def\eeqn{\end{equation}}

\newenvironment{Eqnarray}%
   {\arraycolsep 0.14em\begin{eqnarray}}{\end{eqnarray}}
\def\beqa{\begin{Eqnarray}}
\def\eeqa#1{\label{#1}\end{Eqnarray}}
\def\eeqan{\end{Eqnarray}}

\let\bar=\overbar

\def\lsim{\mathrel{\raise.3ex\hbox{$<$\kern-.75em\lower1ex\hbox{$\sim$}}}}
\def\gsim{\mathrel{\raise.3ex\hbox{$>$\kern-.75em\lower1ex\hbox{$\sim$}}}}

\def\del{\partial}
\def\Dslash{\not{\hbox{\kern-4pt $D$}}}
\def\dslash{\not{\hbox{\kern-2pt $\del$}}}
\def\pslash{\not{\hbox{\kern-2pt $p$}}}
\def\ETmiss{\not{\hbox{\kern-4pt $E$}}_T}

\def\Dlr{\mathrel{\raise1.5ex\hbox{$\leftrightarrow$\kern-1em\lower1.5ex\hbox{$D$}}}}

\def\MSB{{\bar{M \kern -2pt S}}}
\def\msb{{\bar{\scriptsize M \kern -1pt S}}}

\def\drb{{\bar{\scriptsize D \kern -1pt R}}}

\newcommand\snowmass{\begin{center}\rule[-0.2in]{\hsize}{0.01in}\\\rule{\hsize}{0.01in}\\
\vskip 0.1in Submitted to the  Proceedings of the US Community Study\\ 
on the Future of Particle Physics (Snowmass 2021)\\ 
\rule{\hsize}{0.01in}\\\rule[+0.2in]{\hsize}{0.01in} \end{center}}

\begin{document}

\pubblock

\Title{Precision timing for collider-experiment-based calorimetry}

\medskip 

\Author{{\bf Editors:} S. V. Chekanov$^1$, F. Simon$^2$ }
\Address{
$^1$ HEP Division, Argonne National Laboratory, 9700 S.~Cass Avenue, Lemont, IL 60439, USA.}
\Address{$^2$ Max-Planck-Institut für Physik, Föhringer Ring 6, 80805 München, Germany.
}


\Author{ {\bf Contributions from:}}

\Author{V.~Boudry}
\Address{Laboratoire Leprince-Ringuet, CNRS, École polytechnique, Institut Polytechnique de Paris, 91120 Palaiseau, France.}

\Author{W. Chung, M. Nguyen, C.G. Tully} 
\Address{Princeton University, Princeton, New Jersey, USA}

\Author{S.C. Eno, Y. Lai} \Address{University of Maryland, College Park, Maryland, USA}

\Author{A.V. Kotwal}
\Address{Department of Physics, Duke University, Durham, NC 27708, USA}

\Author{S. Ko}
\Address{Department of Physics \& Astronomy, Seoul National University, S. Korea}

\newpage
\Author{P. W. Gorham, R. Prechelt, G. S. Varner}
\Address{Department of Physics \& Astronomy, University of Hawai'i at M\=anoa, USA.}

\Author{S. Lee}
\Address{Department of Physics, Kyungpook National University, S. Korea}

\Author{J.S.H. Lee}
\Address{Department of Physics, University of Seoul, S. Korea}

\Author{I. Laktineh} \Address{Lyon University, IP2I de Lyon 4, rue E. Fermi, 69622, Villeurbanne, France}

\Author{M. T. Lucchini} \Address{INFN \& University of Milano-Bicocca, Italy}

\Author{H. Yoo}
\Address{Department of Physics, Yonsei University, S. Korea}

\Author{C. -H Yeh, S. -S. Yu}\Address{Department of Physics and Center for High Energy and High Field Physics, National Central University, Chung-Li, Taoyuan City 32001, Taiwan}

\Author{R. Zhu}\Address{California Institute of Technology, Pasadena, CA 91125, USA}


\newpage
\medskip

\centerline{
{\Large
{\bf Precision timing for collider-experiment-based calorimetry}
}
}

\begin{Abstract}
\noindent 
In this White Paper for the 2021 Snowmass process, we discuss aspects of precision timing within electromagnetic and hadronic calorimeter systems for high-energy physics collider experiments. Areas of applications include  particle identification, event and object reconstruction, and pileup mitigation. Two different system options are considered, namely cell-level timing capabilities covering the full detector volume, and dedicated timing layers integrated in calorimeter systems. A selection of technologies for the different approaches is also discussed. 
\end{Abstract}

\snowmass

\section{Introduction}

Electromagnetic (ECAL) and hadronic (HCAL) calorimeters are central elements of detectors for high-energy physics (HEP). While their primary purpose is the measurement of the energy of charged and neutral particles and overall event energy, they are also important systems for overall event reconstruction, particle identification and triggering. The physics goals and the experimental conditions at future colliders require technical advances in calorimeter technology to fully exploit the physics potential of these facilities. For future $e^+e^-$ colliders, so-called Higgs Factories, the overall precision of event reconstruction is the main focus, while future hadron colliders at energies and luminosities significantly beyond the HL-LHC impose new challenges in terms of the experimental environment. The detector requirements for future facilities  are given in various reports dedicated to  $e^+e^-$ colliders, such as the Compact Linear Collider (CLIC) \cite{Linssen:1425915}, the International Linear Collider (ILC) \cite{Behnke:2013xla}, the FCC-ee \cite{FCC:2018evy} and Circular Electron Positron Collider (CEPC) \cite{CEPCStudyGroup:2018ghi}, and $pp$ colliders, such as FCC-hh~\cite{FCC:2018vvp} and SppC~\cite{Tang:2015qga}.



The usage of timing information in calorimeter systems has significant potential for further improvements, both in terms of the technology and in terms of reconstruction techniques exploiting this information. 


This paper explores the benefits of precise timing information for calorimetry, and discusses different possible implementations ranging from timing layers with extreme time resolution, timing in larger elements, and volume timing in highly granular calorimeter with moderate time resolution in each cell. 
We expect this discussion  can help shaping the requirements for future calorimeters, which were already
outlined in the CPAD report~\cite{Ahmed:2019sim} that emphasized the need to develop fast calorimetric readouts.


\section{Event and object reconstruction}

The event reconstruction can benefit from the calorimeter timing capacities at several hierarchical levels: timing in cells, in highly granular calorimeters, helps shower reconstruction and energy corrections, timing of individual showers improves particle identification and objects reconstruction, and object timing allows event pile-up mitigation and characterization. 

\subsection{Particle identification}

\begin{figure}
{\includegraphics[width=14cm]{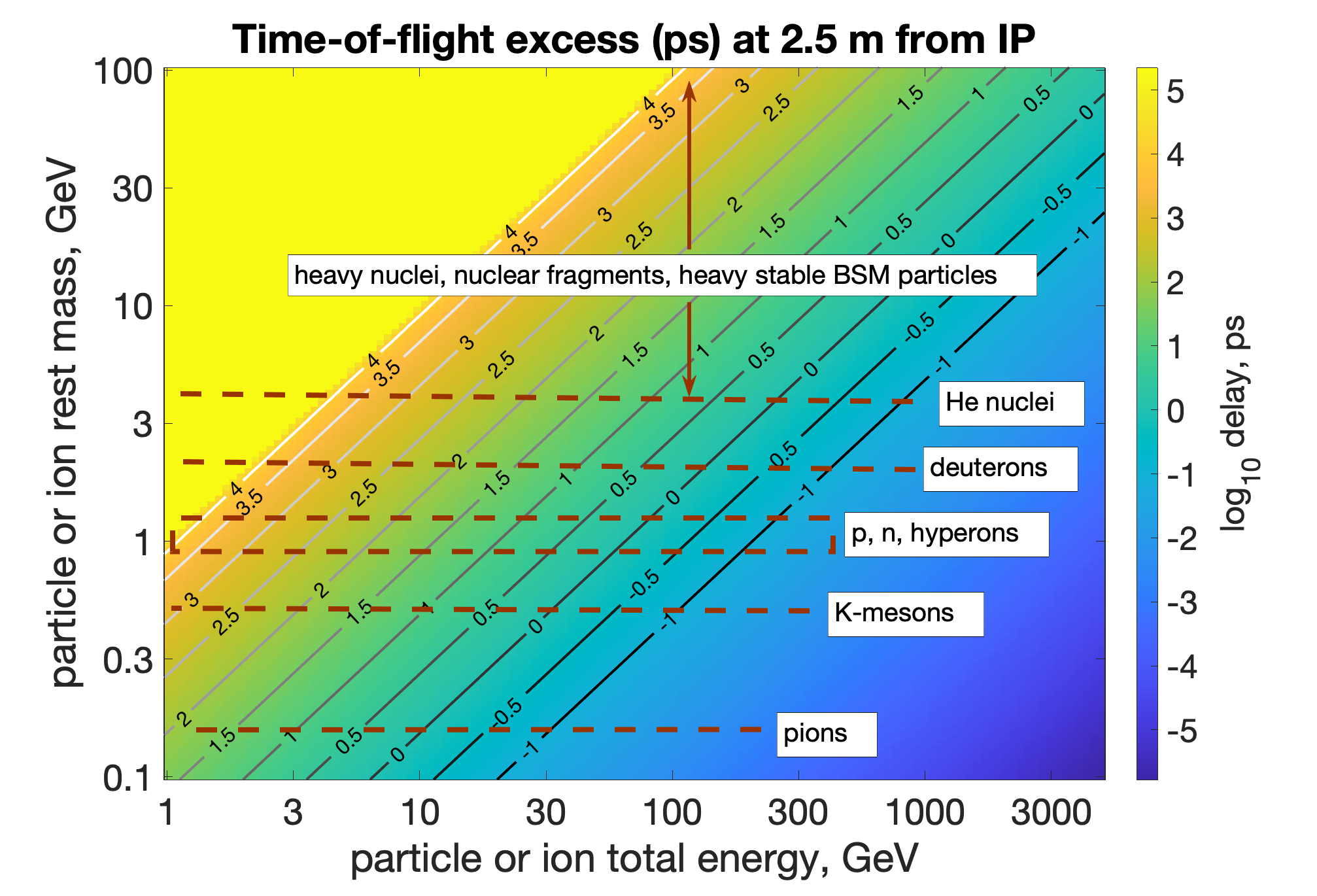}}
{\caption{Time-of-flight for a range of energies and particle or ion masses, for a location typical of a barrel-region calorimeter, 2.5 m from the interaction point.}\label{fig:TOF-2.5m}}
\end{figure}

Timing capabilities of the calorimeter systems of collider detectors open up new possibilities in event and object reconstruction. The concrete possibilities depend on the achieved time resolution and on the technological implementation. Figure~\ref{fig:TOF-2.5m} shows an overview of time-of-flight (TOF) at a typical distance for the barrel region calorimeters, for a range of particle masses, and energies ranging from $\sim 1$~GeV up to several TeV. For timing resolution at the 10~ps level, pions can be resolved up to $\sim 3$~GeV, K-mesons to about 10 GeV, and neutrons and hyperons to several tens of GeV. Heavier nuclei and hypothetical stable BSM particles with be resolved to a high precision at this timing level.

The energy range below 100~GeV, shown in Figure~\ref{fig:TOF-2.5m}, is typical for final-state particle produced in $e^+e^-$ colliders.
For future hadron colliders, such as FCC-pp, average particle energies  will require order-of-magnitude better timing precision than for $e^+e^-$ colliders, i.e. to the picosecond level.
 
With a timing resolution on the order of 10 -- 20~ps for charged hadrons, TOF can be used to identify particles which are heavier than pions. Such a resolution can either be obtained by dedicated timing layers integrated in the electromagnetic calorimeter achieving the required resolution for minimum-ionizing particles, or by a corresponding resolution for hadronic showers provided by the overall calorimeter system.

In order to estimate the separation power between different mass hypotheses as a function of the 
distance between the interaction point and the first layer of the ECAL (or a timing layer to be considered later), one calculate the mass and momentum for which one can achieve a separation significance higher than $3\sigma$ (or p-value$<0.3$\%).
If there are two particles with mass $m$ and a reference (fixed) mass $m_F$, respectively, the $3\sigma$ separation can be
achieved for this condition~\cite{Cerri:2018rkm,2020chekanov}.

\begin{figure}
\begin{center}
   \subfigure[Neutrons] {
   \includegraphics[width=0.45\textwidth]{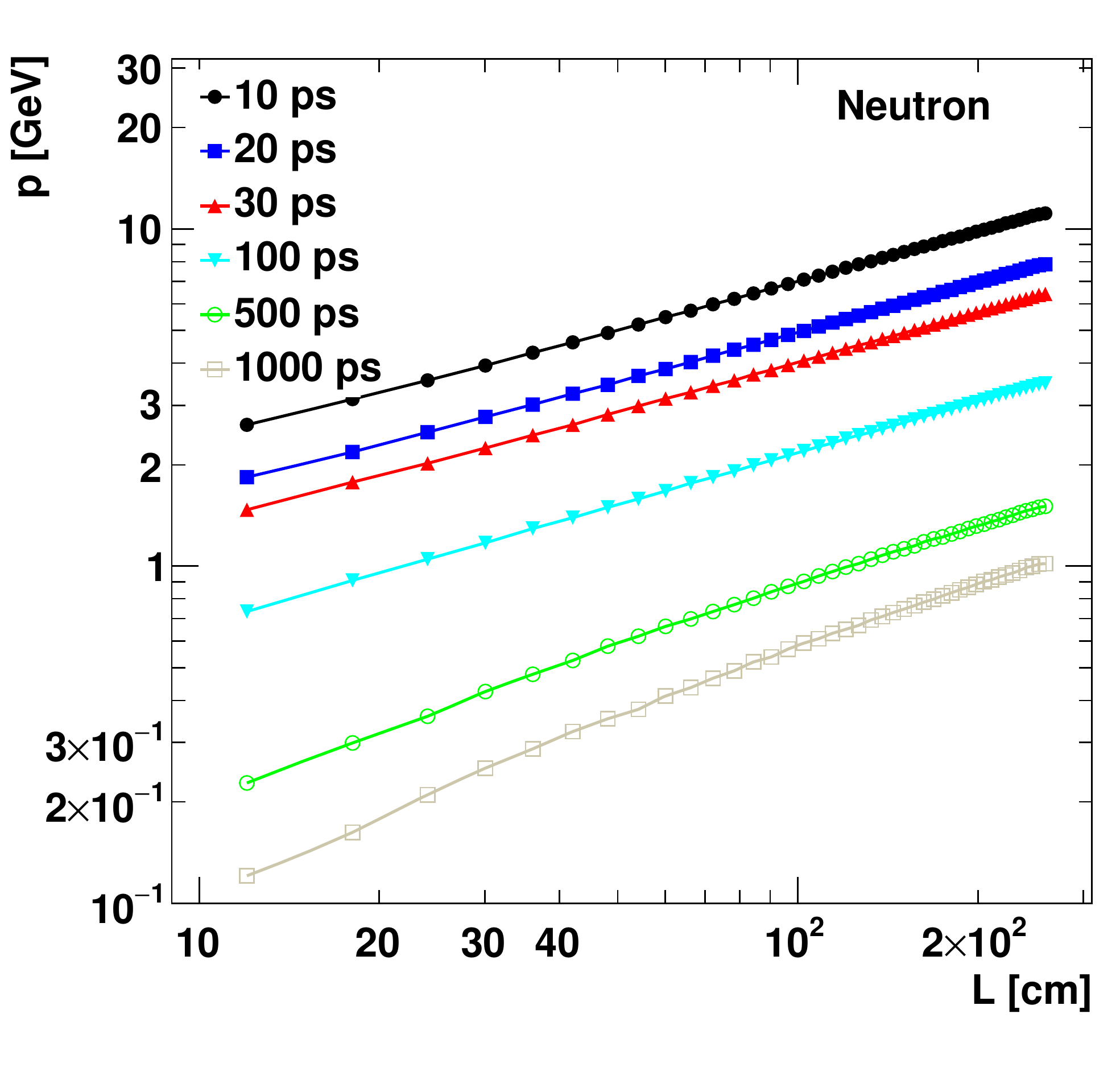}
   }
      \subfigure[$K$-mesons] {
   \includegraphics[width=0.45\textwidth]{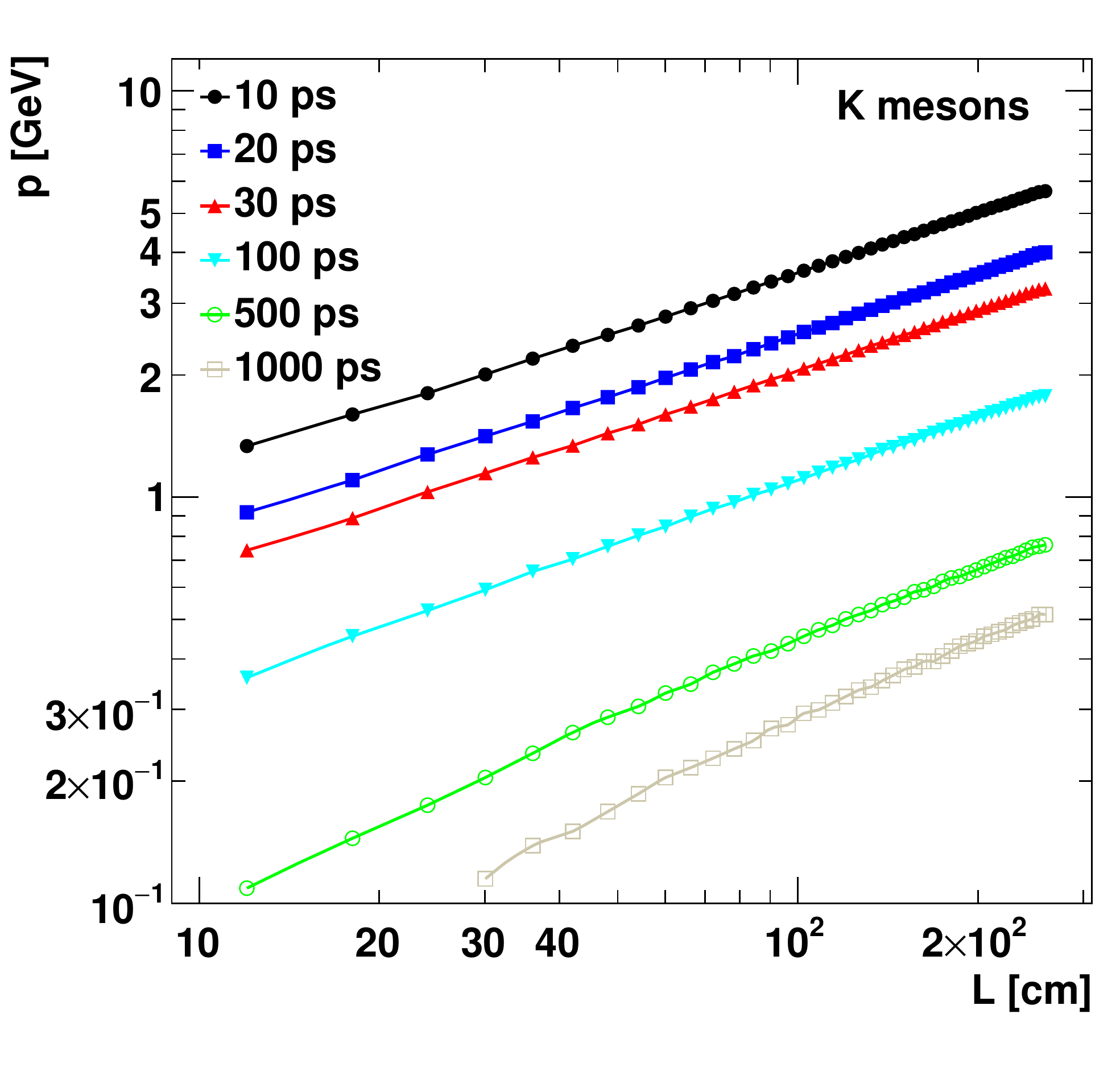}\hfill
   }
\end{center}
\caption{
The $3\, \sigma$ separation from the pion-mass hypothesis for (a) neutrons and (b) kaons as a function of the length of the particle's trajectory $L$ and the momentum $p$. The lines show extrapolated results between the calculations indicated by the symbols.
Reproduced from \cite{2020chekanov}.
}
\label{fig:singleparticles}
\end{figure}

Figure~\ref{fig:singleparticles} shows the $3\sigma$ separation from the pion
mass hypothesis ($m_F=m_{\pi}$) using the procedure discussed  in~\cite{Cerri:2018rkm,2020chekanov} for several values of resolution of the timing layer, ranging from 10~ps to 1~ns. The lines are shown
as a function of the distance $L$ from the interaction point and momentum $p$. For a 20~ps detector and a typical travel
distance $L\sim 1.5-2$~m from the production vertex to the ECAL, neutrons and protons can be separated from the pion hypothesis up to $p \approx 7$~GeV.

According to these studies, separation of kaons from pions can be performed up to 3~GeV.
This momentum range should be sufficient for a reliable particle
identification in a momentum range adequate for some physics studies focused on
single-particle reconstruction (such as B-meson physics).
This can also be used for jets that are dominated
by particles in this momentum range.

\subsection{Identification of long-lived particles}

Searches for new physics beyond the Standard Model (BSM) often predict the existence
of long-lived particles (LLPs) that can give rise to many distinct signatures in calorimeters. 
High-granular calorimeters with precise
timing provide the shower direction and timing information with 
unprecedented precision, enabling us to view them as “tracks” (see \cite{2020LLP}) for a recent review).
Calorimeters with tens-of-picosecond timing  lead to significant benefits
for reconstruction of heavy LLPs. 
For example,  timing layers (or calorimeter cells) with 20~ps resolution lead to almost 100\%
acceptance for large values of dark-pion decay length ($c\tau$) and dark-meson masses \cite{2020chekanov}. This result  is difficult 
to achieve using track-only measurements since only a few outer tracking layers
can be used for LLPs reconstruction. In addition, it has been 
pointed out \cite{2014LLP,2015LLP} that the emerging
jets from the ``dark'' QCD models  may have a significant fraction of neutral particles that
can be measured most efficiently by calorimeter systems.

\subsection{Shower reconstruction and PFA}

Highly granular calorimeters together with Particle Flow Algorithms (PFAs) \cite{THOMSON200925, Sefkow:2015hna} are a widely adopted concept for future collider detectors. With such algorithms, individual final state particles are reconstructed using an optimal combination of tracking and calorimeter information. The overall performance of such algorithms depends on the capability to associate showers in the calorimeters to tracked particles, and on the energy resolution of the calorimeters. Studies have demonstrated that software compensation using the spatial information of the energy density to improve the hadronic energy resolution in the HCAL also significantly improves PFA performance, with contributions both from the improved reconstruction of neutral particles, and a better track-cluster association \cite{Tran:2017tgr}.

Since hadronic showers show a complex time structure, with late components connected to neutron-induced processes, timing on the cell level can have benefits for the spatial reconstruction of hadronic showers. The neutron-driven part of the shower is more diffuse and extended in space than the electromagnetic and relativistic hadronic parts of the shower. A time resolution on the order of a few 100~ps to 1~ns results in a sharper definition of the core part of the shower, and thus potentially in a better separation of different particles in the calorimeter, and improved track-cluster assignment in PFA. A time resolution corresponding to the cell separation times the speed of light, on the order of a few 10~ps or better \footnote{In a given calorimeter layer, cell centers will be separated by their size. Propagation time between layers will involve the interlayer distance. Charge or photon collection inside a uniform cell can also limit the achievable intrinsic (without position correction) resolution to the cell size/2}, may provide additional benefits for pattern recognition, allowing to follow the full space-time evolution of the shower.

First, an early cut on the timing of the cells, along the direction of propagation, displays the “skeleton” of the hadronic showers. Reducing the number of cells in the early stage of the reconstruction dramatically reduces the combinatorial cost of the particle flow algorithms.

\begin{figure}
    \centering
    \includegraphics[width=0.5\textwidth]{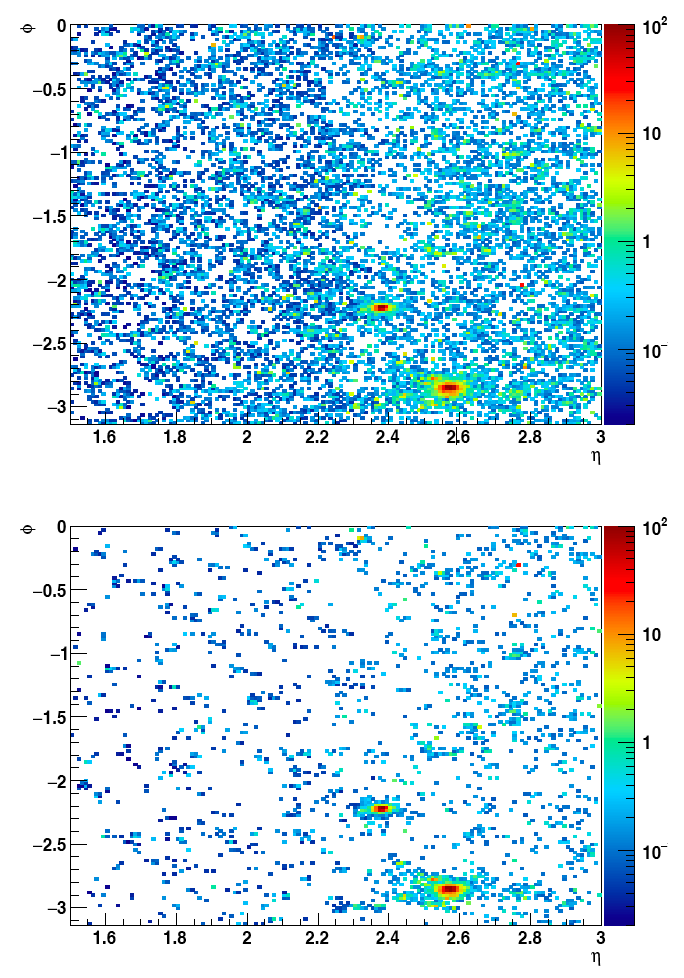}
    \caption{Simulated events of the HGCAL before and after a cut on the time of the particles (from~\cite{CERN-LHCC-2017-023}).}
    \label{fig:ERecoTimingCut}
\end{figure}

The reconstruction of the time-of-flight of a particle at the front face of the calorimeter from the impacted cells is far from simple. It must take into account the correction of the propagation of the components of the shower, which not is linear (e.g. in a magnetic field) and depends on the shower's nature, electromagnetic or hadronic. 
Then, the precision of the cell time roughly scales as the inverse of the signal-to-noise ratio: many cells might bring valuable timing information. As for the measure of the energy, electromagnetic showers are expected to have a better defined time-of-flight than hadronic ones.  Some preliminary studies hint at 10~ps to 50~ps precision for below 30 GeV photons, using 1~ns resolution for at mip level, 3 times more for hadronic showers. This might be improved by having a higher precision in the first  
layers of the ECAL.
Finally, the performances will depend on precise propagation of the signal in the sensors and the associated electronics scheme and the possible corrections. Those need to be properly defined and simulated before drawing conclusions.


\begin{figure}
    \centering
    \includegraphics[width=0.5\textwidth]{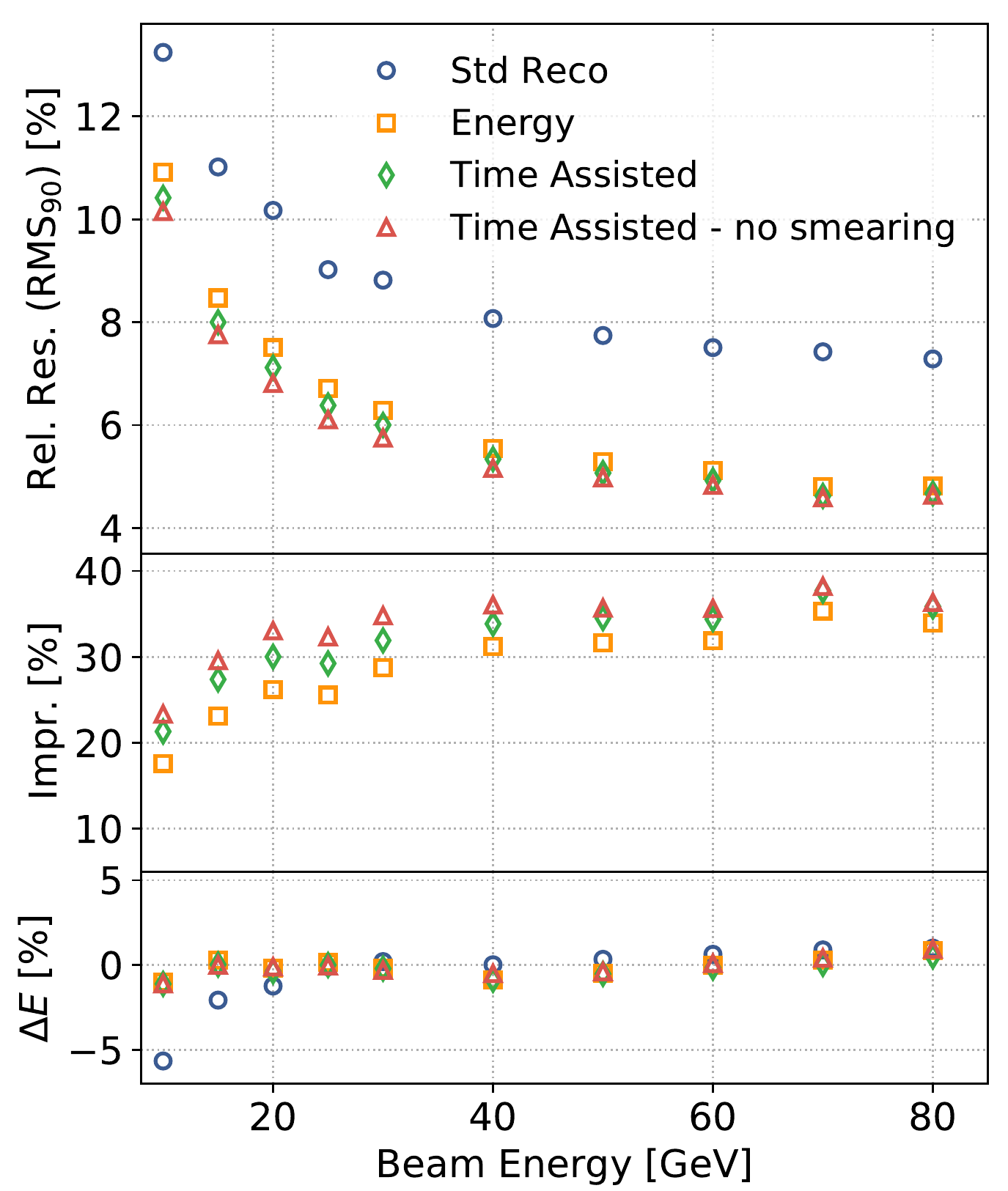}
    \caption{Simulated single-hadron energy resolution for a highly granular SiPM-on-tile calorimeter closely modelled on the CALICE AHCAL \cite{Sefkow:2018rhp} with local software compensation using energy density alone, as well as timing with 1~ns resolution and with a perfect time resolution (corresponding to a few 100~ps, in practice). Figure taken from \cite{Graf:2022lwa}.}
    \label{fig:ERecoWithTiming}
\end{figure}

Along the same lines, highly granular time information of hadronic showers can also be used in software compensation techniques as an additional dimension, with the potential for further improvement of the energy resolution. Simulation studies performed in the context of the CALICE SiPM-on-tile analog hadron calorimeter show that a cell-by-cell time resolution on the ns level for MIP-equivalent energy depositions results in an increase of the improvement of the energy resolution by 10\% to 15\% compared to a purely energy-density-based local method \cite{Graf:2022lwa}. The gain provided by timing can be approximately doubled with significant better time resolution, as illustrated in Figure \ref{fig:ERecoWithTiming}. The study also illustrates the significant degree of correlation between energy-density-based and time-based shower observables, limiting the additional gain provided by timing. Both are sensitive to the share of shower activity between electromagnetic and hadronic and hadronic components. The local energy density is primarily sensitive to the electromagnetic fraction, while late shower components accessible via timing provide sensitivity to neutron-induced processes, and thus track hadronic shower activity.  

Measurement of timing along tracks in the calorimeters can also provide the beta and $dE/dx$ of long travelling particle inside the shower, helping their identification, hence energy reconstruction, or correction of their leakage.     

More complex reconstruction algorithms can potentially make even better use of the multi-dimensional information provided by highly granular calorimeters with timing. First promising results have been achieved with convolutional and graph neural networks, with a performance increasing with time resolution \cite{Akchurin:2021afn}.

\begin{figure}
\begin{center}

\subfigure[] {
   \includegraphics[width=0.45\textwidth]{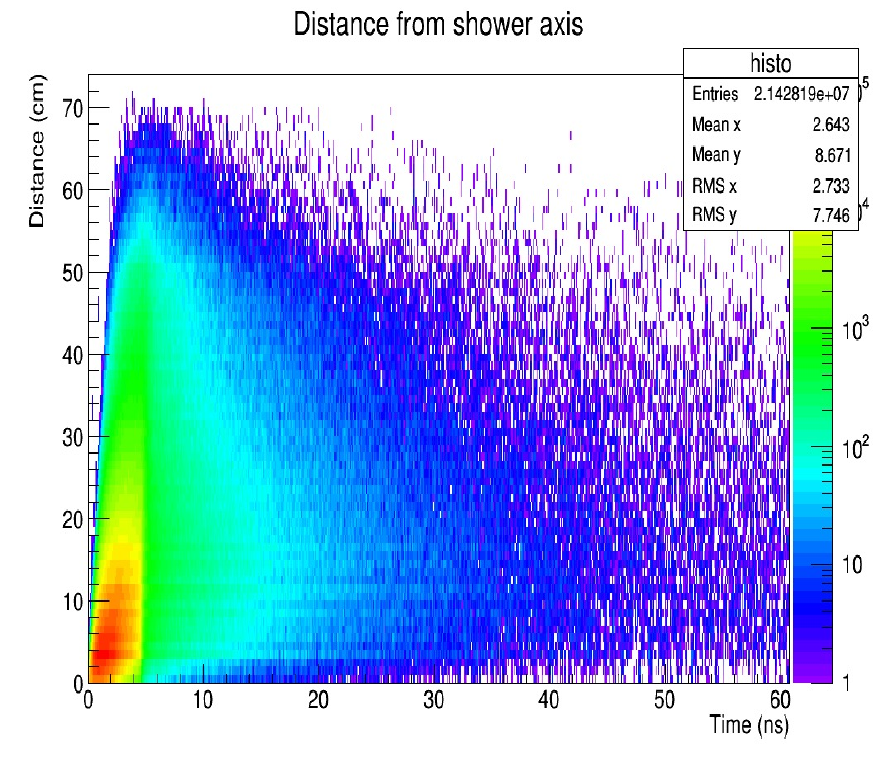}
   }
   \subfigure[] {
   \includegraphics[width=0.45\textwidth]{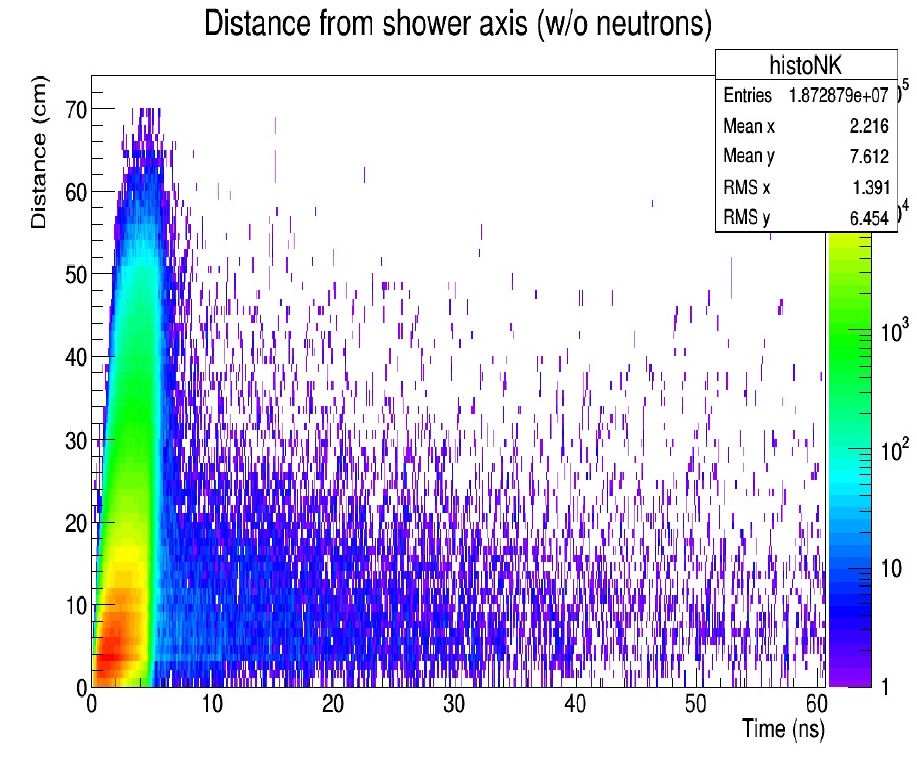}
   }
\end{center}
\caption{Hit distribution versus the distance from the shower axis and time for all hits (left) and with the  hits without those originating from neutrons (right), simulated for the RPC-based CALICE SDHCAL.}
\label{fig:semdig}
\end{figure}

The sensitivity to different parts of the shower also strongly depends on the type of active medium. While in particular organic scintillators with their large hydrogen content are sensitive to MeV-scale neutrons, gaseous detectors, such as RPCs, Micromegas or GEMs, provide significantly less sensitivity to neutrons, resulting in less sensitivity to the wider-spread neutron component of hadronic showers. Timing on the sub-ns level further helps to increase the separation of close-by showers. This is illustrated with simulations performed in the context of the RPC-based highly granular CALICE semi-digital hadron calorimeter (SDHCAL) \cite{Baulieu:2015pfa} in Figure~\ref{fig:semdig}. The figure shows the hit distribution versus the distance from the shower axis and the time of all hits (left) and the hits without those originating from neutrons (right).
The neutron-induced hits introduce fluctuations on the event-by-event basis that deteriorate the resolution of the energy reconstruction. They also increase, by their geometrical distribution, the confusion between two nearby showers and decrease thereby the possibility to separate them. With sub-ns-level time resolution, as provided by RPCs with a single gas gap, late hits originating primarily from neutrons can be included in a first step of the shower reconstruction. In subsequent steps, the time and geometrical information of these hits can then be used for an improved assignment to the correct shower. In addition, techniques for energy reconstruction as the one outlined above for scintillator-based calorimeters, are also applicable here. 

Beyond single shower reconstruction, the efficient separation of particle showers in higher-density environments is important for PFA. In the absence of time information, the separation between two hadrons impinging on the SDHCAL with a distance less than 10 cm decreases significantly, reaching only 60\% at a separation distance of 5 cm. The connection of hits belonging to the same shower is based on their position in the different reconstruction algorithms. Close-by hits in space are associated to each other and then assigned to the same shower even if they occur at two different times. Comparable behaviour is also observed for other detection media. 

Significant further improvement is expected with a time resolution that is better than the one imposed by the causality relation linking two neighboring hits. This requires a time resolution comparable to the granularity of the calorimeter, multiplied by the speed of light, corresponding to a few 10 to 100 ps. Different technologies exist that are in principle capable of delivering such a performance on the cell level, as briefly outlined in Section \ref{sec:technology:volume}.

\subsection{Longitudinal shower reconstruction for fiber sampling calorimeters}

The time information can also be used to reconstruct shower shape for longitudinally unsegmented fiber sampling calorimeters like the dual-readout calorimeter \cite{RevModPhys.90.025002}, where optical fibers are inserted longitudinally then attached to photodetectors such as SiPM at the rear end. As the energy deposit closer to the readout requires a much shorter propagation time than the more distant one, the difference between the speed of incident high-energy particles and that of emitted photons inside optical fibers provides us a tool to estimate energy density shape along the longitudinal axis.

\begin{figure}
\begin{center}
   \subfigure[Depth] {
   \includegraphics[width=0.45\textwidth]{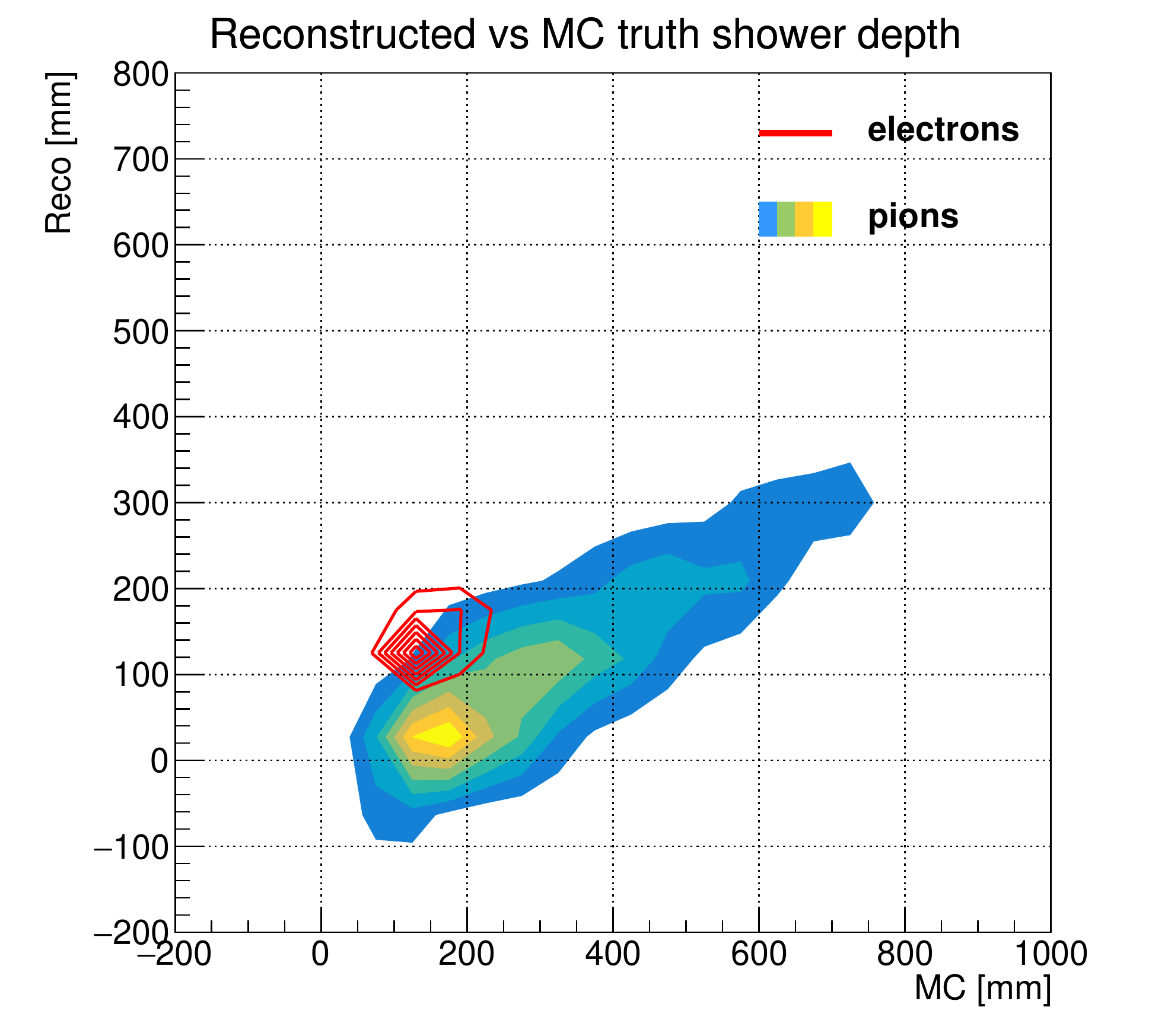}
   }
      \subfigure[Length] {
   \includegraphics[width=0.45\textwidth]{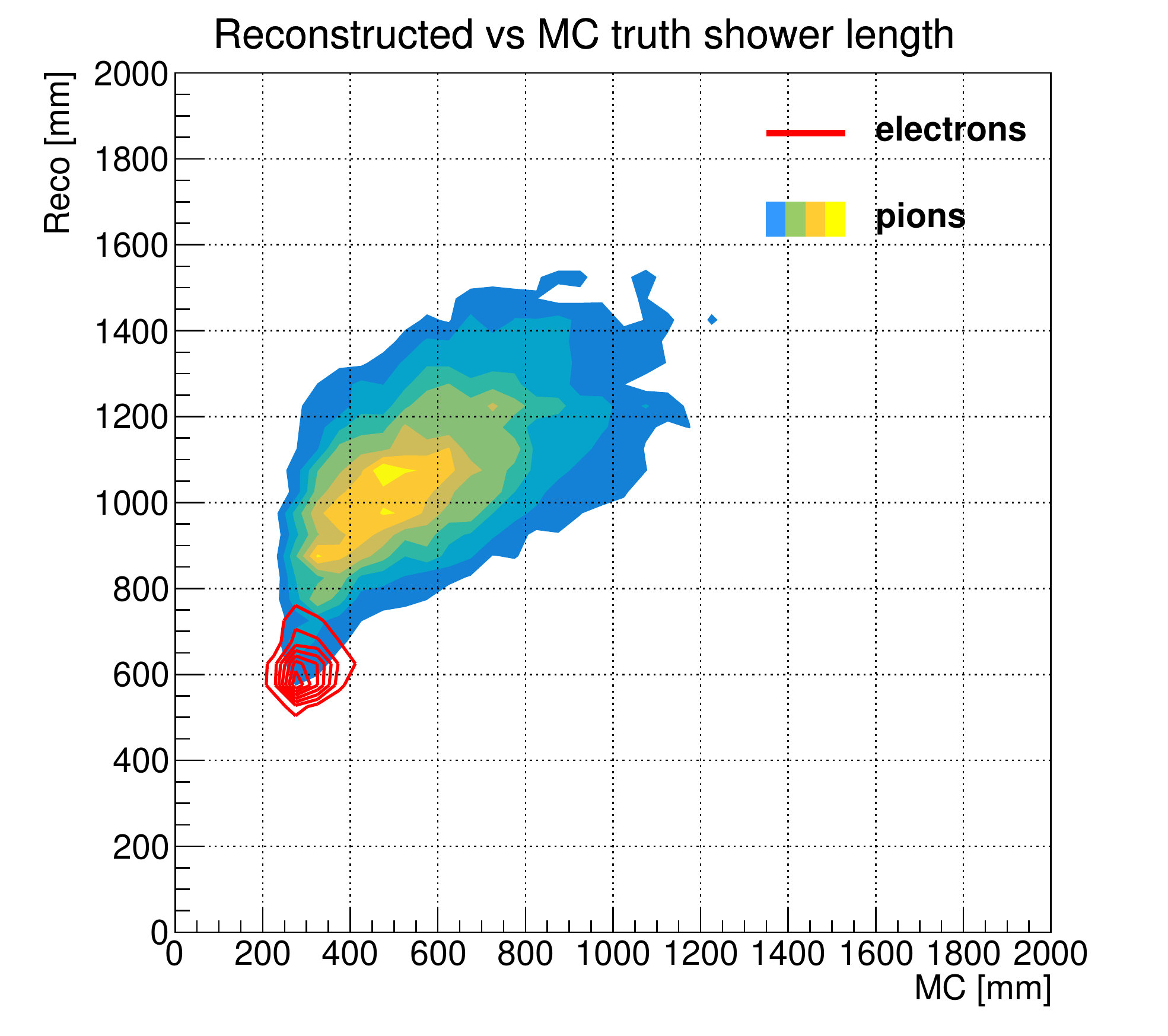}\hfill
   }
\end{center}
\caption{
Correlation between the reconstructed and MC truth shower depth (a) and length (b) for 20 GeV electrons (lined contour) and pions (filled contour) in the dual-readout fiber sampling calorimeter \cite{RevModPhys.90.025002} using the timing distribution of detected photons in GEANT4 simulation.
}
\label{fig:showershapeDRC}
\end{figure}

The timing distribution can be interpreted as the energy density shape by deconvoluting exponentiating components from detected signals, which can be caused by the response of photodetectors to the unit pulse and the scintillation decay. We measured the depth of a shower maximum and the length of a shower, defined as a distance between the furthest two points exceeding 10\% of energy density at the peak, in an event after applying the signal processing. Figure \ref{fig:showershapeDRC} shows the correlation between the reconstructed and MC truth shower depth and length for 20 GeV electrons and pions events using GEANT4 simulation assuming a 100 ps sampling rate, without rescaling interpreted time observables.

A significant difference in the shower length between electrons and pions well describes a feature of event-by-event fluctuation of hadronic showers, illustrating that the timing is utterly useful for particle identification. Moreover, a certain correlation between the reconstructed and MC truth observables demonstrates the possibility of using longitudinal shower shapes even for the longitudinally unsegmented calorimeters by exploiting time observables.

\subsection{Jet substructure reconstruction}

\begin{figure}
\begin{center}
   \includegraphics[width=0.9\textwidth]{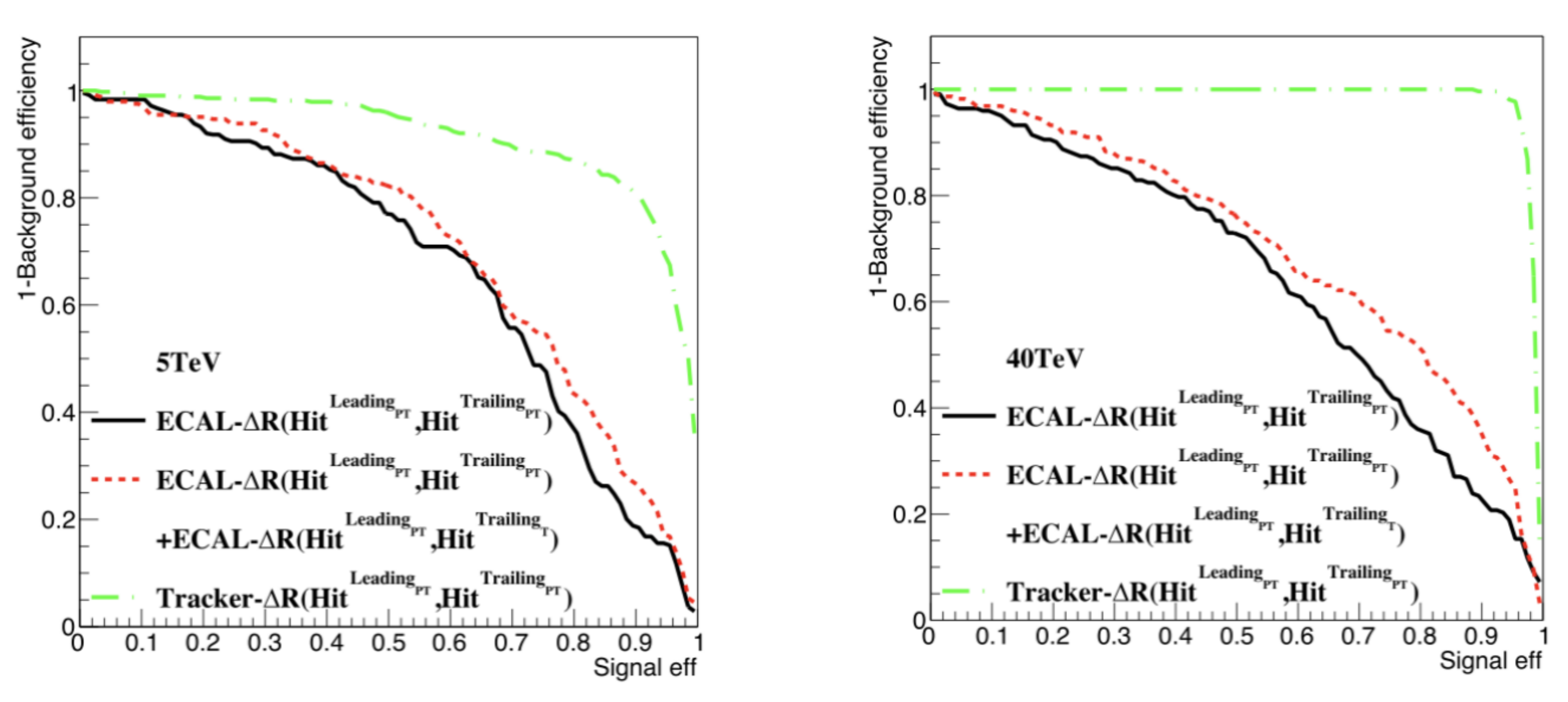}
\end{center}
\caption{
The Receiver Operator Characteristic (ROC) 
curves for $Z'$  mass of 5 TeV (left) and 40 TeV (right), obtained using the following 
variables: (i) $\Delta R$ between the leading-$p_T$ and the trailing-$p_T$ particles (black), and (ii) adding the information of $\Delta R$ between the leading-$p_T$  and trailing-$T$  particles (red).
The ROC obtained using the tracker-information only is also shown for comparison (green). 
All the physical variables, including 
$\Delta R$ and the ranking in $p_T$ and $T$, 
are obtained from the reconstruction-level 
information. 
}
\label{fig:roc_jets}
\end{figure}

Another area where high-precision timing is  potentially important is the reconstruction of jet substructure for
resolving highly-boosted objects. Some theoretical ideas with the support of truth-level Monte Carlo simulations indicate that 
a boosted object tagger incorporating timing information can be  complementary to traditional taggers, and may help discriminating
signals against background events~\cite{2021jettime}. In the context of the Snowmass 21 study, 
there was an attempt to understand the time structure of jets
using a full Geant4 simulation ~\cite{Allison2016186}  of a generic future detector~\cite{2017fcc,2019granularity}   for a 100~TeV proton-proton collider.

In these studies, a hypothetical heavy $Z'$ gauge boson, postulated in extensions of the standard
model, is simulated with the  masses of 5, 10, 20, and 40~TeV. The $Z'$ bosons are forced
to decay to two light-flavor jets $q\bar{q}$ to model the background and to $WW$ pairs  where
the W bosons decay hadronically to model the boosted $W$ boson ''signals''. It was explored
how the background rejection can be achieved by including the timing information, in addition to the
measurement of particle transverse momenta.

Particles and calorimeter hits of jets from $Z'$ were  ranked by their
transverse momenta $p_T$  and by their TOF. Then the signal to background separation was calculated for: (i) exploiting only
the $p_T$ of the trailing-$p_T$ particles, (ii) adding the TOF of the trailing-$T$ (slowest) particles at the
truth-level, and (iii) adding the measured TOF of the trailing-$T$ particles.

Figure~\ref{fig:roc_jets} shows the the Receiver Operator Characteristic (ROC) 
curves for $Z'$  mass at 5 TeV (left) and 40 TeV (right), obtained using the following 
variables at the reconstruction level: (i) $\Delta R$ between the leading-$p_T$ and the trailing-$p_T$ particles (black), and (ii) adding the information of
$\Delta R$ between the leading-$p_T$  and trailing-$T$  particles (red). 
The results show that the timing information does improve the background rejection for 
highly-boosted jets from $Z'$ mass above 20~TeV. Below this this, no significant improvements have been seen.
However, more studies in this direction will require to confirm this observation.

\subsection{Pileup mitigation}

In high-energy proton collisions, the biggest challenges in detector design arise from a large number of pile-up events (interactions per beam crossing). For the FCC-hh machine, about $O(1000)$ effective pileup events are expected.
For low momentum particles, a 2D vertexing with an extreme timing resolution 
5--10~ps per track is essential \cite{Drasal:20181K}.
For example, 90\% assigned tracks in the central region can be achieved with ~5--10 ps timing cuts \cite{FCC:2018vvp}, that will keep the effective pile-up below one  per bunch crossing within the central region. A calorimeter with a tens-of picosecond resolution should bring  additional improvements using both charged and neutral particles.  

\subsection{The picosecond/sub-picosecond frontier.}
\label{microwaveCherenkov}

A timing resolution of 10~ps has been used as a benchmark above, based on recent advances in ultra-fast silicon detectors (UFSD), which have recently achieved a timing precision of $16$~ps~\cite{CARTIGLIA201783}, demonstrated through combining signals from a three-element detector ensemble. These devices, with use low-gain avalanche detectors tailored specifically to improve their timing characteristics, are plausible candidates to achieve 10~ps in the near future. The one difficult challenge currently facing these devices is their loss of gain at high radiation fluences; recent work has shown substantial gain damage at neutron equivalent fluences of $\sim10^{15}$~neq/cm$^2$~\cite{LGAD21}, more than an order of magnitude below the exposure expected at the HL-LHC or the Future Circular Hadron Collider~\cite{Benedikt:2018csr}.
Thus the challenges at the timing frontier are two-fold: to push the time resolution down to the 10~ps level and below, and at the same time to develop radiation-hard devices and systems as colliders push into regions of radiation fluences that are far above current levels. 

\begin{figure}
\begin{center}
   \includegraphics[width=0.5\textwidth]{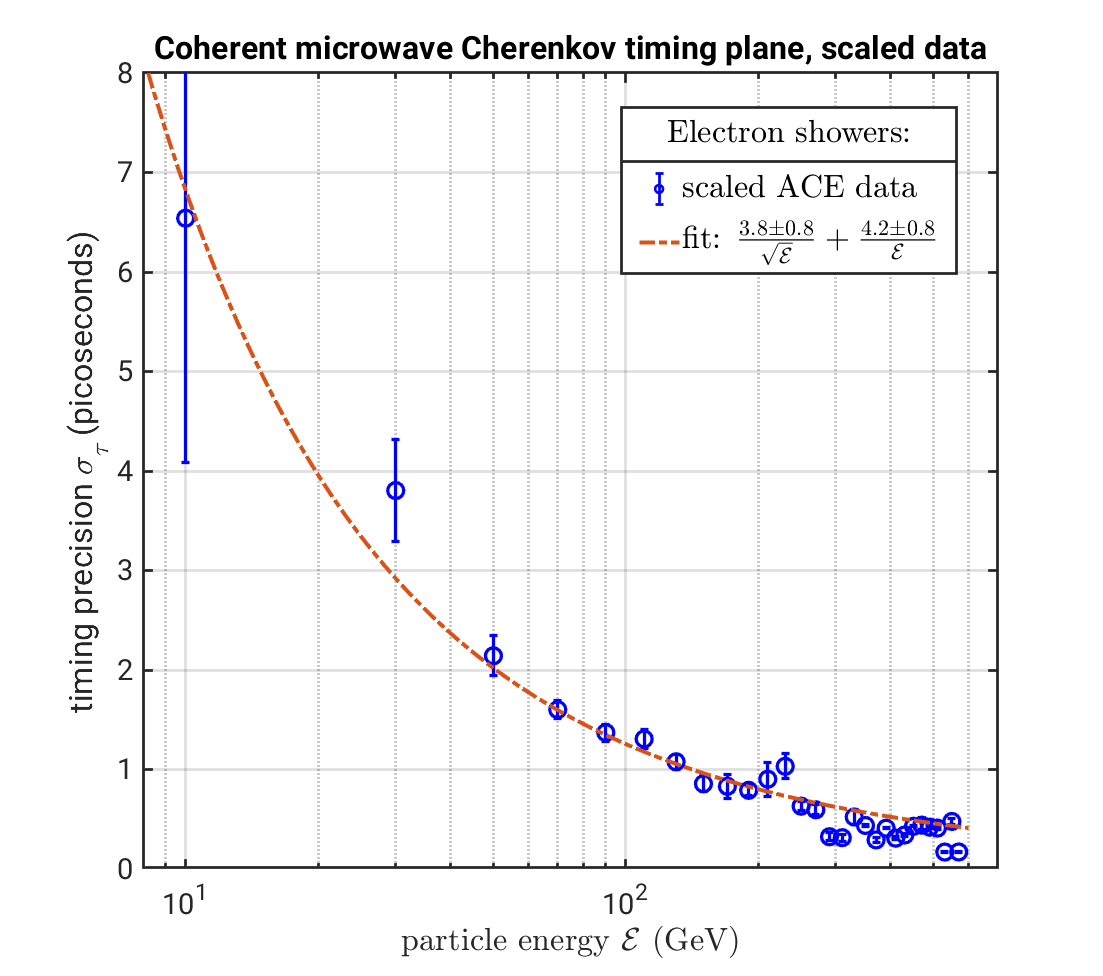}\includegraphics[width=0.5\textwidth]{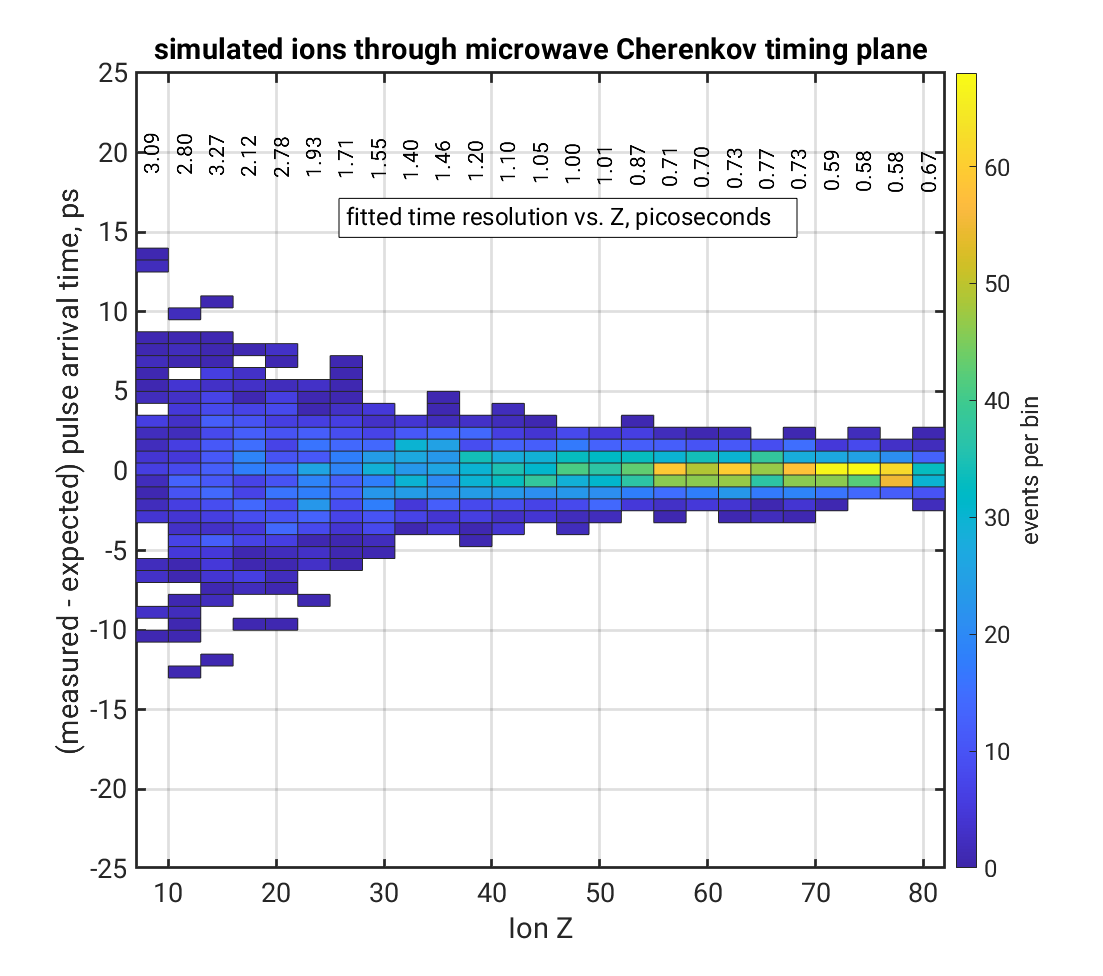}
\end{center}
\caption{
Two examples of the potential of coherent microwave Cherenkov timing, based on results from the ACE collaboration~\cite{gorham2022picosecond}. Left: Data from a 2018 beam test of electron showers with a single element, scaled to a full-timing layer with liquid argon cooling. This estimate assumes a pre-shower timing layer with $\sim 3.5X_0$ upstream of it. Right: Timing of heavy ions, via coherent Cherenkov from their $Z$. Coherence in this case is significantly higher since nuclear charge is complete unresolved at microwave frequencies.
}
\label{fig:psTiming}
\end{figure}

One new technology under development with a view toward the FCC is coherent microwave Cherenkov detection, which has now demonstrated $\sim 2-3$~ps timing of electromagnetic showers using dielectric-loaded rectangular wave-guide elements~\cite{PhysRevAccelBeams.21.072901}. While the UFSD described above may achieve 10~ps timing for a tracking detector, no other technology has achieved picosecond timing for showers, which are the natural domain of calorimeters. Microwave signals, multi-GHz bandwidth and center frequencies, have rise times an order of magnitude faster than solid-state detectors. They have very high dynamic range, $>10^4$, and inherent radiation hardness. The current drawback of this method is the relatively high least-count energy, several tens of GeV, due to the fundamental limits of thermal noise, but for colliders at 100~TeV center-of-momentum energy, this limitation may have much less importance, as the mean momentum of jets and collision products shifts to much higher energies and to the forward direction~\cite{gorham2022picosecond}. 

Estimates based on recent beam test data indicate that $\leq 1$ picosecond timing is achievable for both electromagnetic and hadronic showers over a large range of the collision products at the FCC-hh~\cite{gorham2022picosecond}. Figure~\ref{fig:psTiming}, adapted from reference~\cite{gorham2022picosecond}, shows an example of timing precision using this emerging methodology. On the left, data from a 2018 beam test of electromagnetic showers in a single dielectric-loaded wave-guide element is scaled to a fully optimized multi-element timing layer, showing a transition to timing precision of several picoseconds in the 10-100~GeV range, and drops to the subpicosecond regime above 100~GeV. On the right in Fig.~\ref{fig:psTiming}, simulation results for heavy ions timing using this technique are shown. In this case, the coherence of the microwave emission is significantly higher because the nuclear charge is completely unresolved at microwave frequencies, and thus showering of the ion is not required to detect and time it, even to sub-picosecond levels for the heavier ions.

\section{System options}

When considering technology for timing detection, one should take into account several factors, such as the physics potential and the price for instrumentation. 

Below we will discuss several possible options. 

\subsection{Volume timing}

Under ``volume timing''  we understand  timing information to be collected by all active calorimeter cells. The implementation in a highly granular calorimeter enables a full five-dimensional reconstruction of shower activity in the detector, with corresponding benefits for pattern recognition, spatial shower reconstruction and separation and energy measurement as outlined above. From a physics perspective, this technology represents the best possible choice since all information from readout cells can be included in physics analysis. At the same time, the very large number of channels in such designs may require compromises on the timing resolution for cost or technology reasons. Besides, good “classical” calorimetric and timing performances might be 
contradictory: a good energy resolution implies the collection of a large signal, hence cell volumes, while the timing might be limited by the fluctuations linked to the volumes needed (see~\cite{riegler_time_2017} for the case of silicon sensors). A possible solution might lay in the development of fully digital calorimetry: some work has started (CALICE SDHCAL, ALICE FOCAL) but will still require significant efforts on the technology.

An alternative option is to integrate high-precision timing in several dedicated
calorimeter cells separated by some distance in the transverse (to the beam) direction, providing timing only for a subset of all detector cells. This may enable more aggressive time resolution goals in a fraction of the overall detector volume while still respecting cost and technology constraints. 
If such cells are located in the first (last) layers of the ECAL, this option will be similar to the ``timing'' layer choice to be discussed below.


\subsection{Timing layers}
\label{ref:tl}

Building a full-scale calorimeter system with the primary sensing elements with a tens-of-picosecond
resolution for all cells can be challenging. 
As a  possible alternative, dedicated detectors can be installed on the front of the ECAL.  An LYSO+SiPM  timing layer aiming at 30 ps timing resolution is currently under construction by the CMS experiment for the HL-LHC \cite{Butler:2019rpu}.
The potential for such timing layers (TLs) was studied in the context of future collider experiments for example in \cite{2020chekanov}.

A similar idea is discussed for the Segmented Crystal Electromagnetic Precision Calorimeter (SCEPCal) \cite{Eno2020} with dual readout.
The proposed detector consists of two thin
layers with the capability of measuring single MIPs with a time resolution of about 20~ps.
Each layer is made of inorganic scintillator square fibers close to each other.

\begin{figure}
\floatbox[{\capbeside\thisfloatsetup{capbesideposition={right,center},capbesidewidth=5cm}}]{figure}[\FBwidth]
{\caption{Example positions of timing layers (TLs) for a generic detector. The timing detectors  enclose the
electromagnetic calorimeter, allowing  a reliable calculation of the  MIP signals with a timing resolution of the order of 10~ps.}\label{fig:eff_rad}}
{\includegraphics[width=8cm]{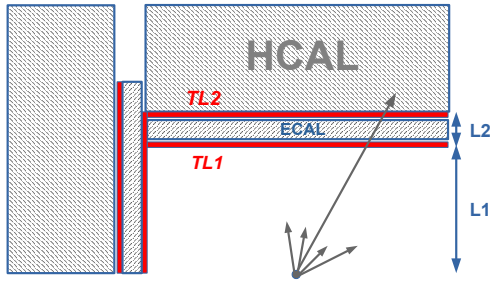}}
\end{figure}

The usage of detectors with timing capabilities on the front of the ECAL is not new. Unlike the calorimeter ``volume'' systems, the timing detectors can be optimized for precise time measurements using different types of  technologies. Such detectors can also be optimized for granularity, and ultimately, for the price per channel.
One new study conducted during  Snowmass21 was focused on the 
verification of the usage of several timing layers, one before the ECAL, 
and the second layer - after the ECAL.
The second timing layer can be used to measure the TOF between TL2 and TL1 in the identification of stable massive particles without a known production vertex, correlate the hits with the first layer, and thus provide directionality
of the hits. Finally, it can be used for redundancy for the calculation of TOF.
A schematic view of two timing layers for a generic detector is shown in Fig~\ref{fig:eff_rad}.  It was verified that  a typical time difference between TL2 and TL1 (which is approximated by the difference
between the last and first ECAL layer) is sufficient  \cite{2020chekanov}  for identification of low-mass particles below the GeV-scale in momentum assuming the tens-of picosecond resolution of both timing layers.

\section{Possible technologies}
\label{sec:technology}

In this section we will consider several possible technology choices for the implementation of precision timing in calorimeters with resolutions in the range of a few tens of picosecond. We divided such technologies
into two categories to be discussed below.  

\subsection{Technologies for timing layers}

Here we will discuss possible technologies that can be used for precision time stamping of charged particles entering electromagnetic calorimeters
(or the hadronic calorimeters in the case of the designs with several
timing layers discussed in Sect~\ref{ref:tl}).
In addition to the excellent time resolution,
small radiation lengths,  small thickness and  radiation tolerance are the
main technology requirements for timing layers.
Space for the cooling, readout and other services that are required for the 
timing layers can be allocated in the front and the back of the  ECAL/HCAL sections.

\begin{itemize}
    \item Low-Gain Avalanche Detectors (LGADs) developed for High Luminosity LHC experiments have demonstrated precise time resolution  of about 30~ps.  The position resolution of 1~mm is sufficient for most physics cases for calorimeter measurements.  
    
    \item
    Ultra-fast silicon monolithic sensors with integrated electronic readout using the CMOS technology, in an effort to achieve ~20-10 picosecond timing resolution. They are 
     also  expected to significantly reduce costs while maintaining radiation hardness. An example of this sensor, Depleted Monolithic Active Pixel Sensor (DMAPS),
     is discussed in \cite{Degerli_2020}
     
     \item
     Micro-channel plate (MCP) detectors enable the detection of single ionizing particles to a precision of a few ps \cite{Ronzhin:2014lqa,BORTFELDT2020163592}. For example, Micro-Channel-Plate Photomultiplier Tubes (MCP-PMT), R3809U-50, by Hamamatsu,
    indicate a time resolution better than 10~ps \cite{BORTFELDT2020163592}. 
    
    \item
    A two-stage  Micromegas detector coupled to a Cherenkov radiator equipped with a photocathode can lead to  timing resolution for charged particles significantly below 100~ps \cite{Bortfeldt:2017daz}.
    
    \item
    Sampling calorimeters based on a Lutetium-yttrium oxyorthosilicate  (LYSO) crystals were demonstrated to  have  a time resolution  in the range of a few tens of picosecond  \cite{7581887}.  Such scintillating crystals coupled to a SiPM represents a flexible option that has been proposed for the Segmented Crystal Electromagnetic Precision Calorimeter \cite{Eno2020}.
    
    \item
    Deep diffused avalanche photodiodes which were demonstrated \cite{CENTISVIGNALI2020162405}  to achieve the time resolution of about 40~ps.
     
   \item
   Coherent microwave Cherenkov detectors, as noted in section~\ref{microwaveCherenkov} above, have demonstrated $\leq 3$~ps timing of energetic electromagnetic showers~\cite{PhysRevAccelBeams.21.072901}. Coupled electromagnetic + GEANT4 models of the process indicate that $0.3-1$~ps timing of electromagnetic and hadronic showers is already achievable for typical FCC-hh energies~\cite{gorham2022picosecond}.

\end{itemize}

\subsection{Technologies for volume timing}
\label{sec:technology:volume}

The technology for the ``volume'' timing should take into account the fact that the same 
active material should be used for the energy and timing measurements.
This list encapsulates the technologies that can be most appropriate for calorimeter cells where
both the time-of-arrival and energy of the particle are measured in the same active detector element.

\begin{itemize}
  
    \item Silicon tiles in different implementations. Achieving a few 10 ps for single particles requires sensors with intrinsic gain, such as LGADs, while precise timing for high-amplitude signals is also possible with conventional sensors. One example of such an application is the CMS HGCAL, where the timing information from many individual cells is used for a precise overall determination of the time of physics objects \cite{Lobanov:2020hhr}.
   
    \item Plastic scintillator tiles or strips with SiPM readout. SiPMs are intrinsically capable of a few 10 ps level resolution, a key factor is the scintillator response and the light collection, and the light yield, which drives the stochastic aspect of the time resolution. For the CALICE SiPM-on-Tile analog hadron calorimeter, sub-ns time resolution is achieved for minimum-ionizing particles on the cell level \cite{Emberger:2021lsz}. Ongoing studies show that this resolution is driven by the number of detected photons, which can be increased by increased sensor size, and the emission characteristics and light collection effects in the scintillator. 
    
    \item Resistive plate chambers, in particular multi-gap RPCs provide a high time resolution and can be used to efficiently cover large active areas for digital and semi-digital hadronic calorimeters. Detectors with four gas gaps with less than 300 $\mu$m each can provide sub-100 ps time resolution, which can be exploited on the system level with suitable electronics (see, for example, \cite{app11010111}). A small prototype based on 4-gap RPCs and PETIROC ASICs is currently in development in the framework of the SDHCAL project within CALICE, with a larger prototype, referred to as T-SDHCAL, as a longer-term perspective in case of adequate funding. 
    
    \item Highly granular crystal-based detectors, using a highly segmented readout based on small scintillating crystals or other high-density scintillating materials. As for plastic-scintillator-based highly-granular sampling calorimeters key factors for the time resolution are the time characteristics of the scintillation emission and the light yield. The higher density of crystals and the correspondingly larger energy loss can, with a suitably fast scintillator, result in faster timing than for common plastic scintillators in otherwise similar geometries. \\
    One example are ultrafast heavy crystals with sub-nanosecond decay time, which would help to break the ps timing barrier for precision timing. Inorganic scintillators with core valence transition features with its energy gap between the valence band and the uppermost core band less than the fundamental bandgap, allowing an ultrafast decay time. Decay time of  0.5 nanosecond is observed for barium fluoride (BaF$_2$) crystals. Ultrafast crystals with mass production capability are discussed in \cite{bib:ultrafast}.

    \item Digital Silicon PhotoMultipliers (dSiPM), based on arrays of digitally combined and managed Single Photon Avalanche Diodes (SPAD), might provide better than 20~ps time resolution for single mips \cite{therrien_energy_2018, riegler_time_2021}. With 3D integration, they might have consumption and integration compatible with large calorimeters.
    
\end{itemize}

Common to all volume timing technologies is the need for electronics that support the required time resolution while also satisfying the constraints on power consumption associated with highly integrated systems with extreme channel counts. R\&D in this area is critical for the further development of these calorimeter concepts.

\section{Conclusion}

High-precision timing is emerging as a key capability of calorimeter systems at future high-energy-physics collider experiments, expanding capabilities for particle identification, energy measurement, shower reconstruction and particle-flow event reconstruction, as well as pile-up mitigation and background rejection. Two different conceptual approaches exist, with dedicated timing layers providing high-precision measurements within calorimeter systems at selected longitudinal positions, and volume timing delivering uniform, often somewhat less aggressive, time resolution for all detector cells in the calorimeter. A wide range of technologies is being explored in this context, ranging from silicon-based sensors to organic and inorganic scintillators coupled to different types of photon sensors and gaseous detectors. Calorimeters with such capabilities, providing time resolutions in the range of a few tens of picoseconds are expected to be an important step towards accurate reconstruction of events at future colliders.

This work has been performed during  the Particle Physics Community Planning Exercise, Snowmass 2021, that was organized by the Division of Particles and Fields  of the American Physical Society.

\section*{Acknowledgments}
The submitted manuscript has been created by UChicago Argonne, LLC, Operator of Argonne National Laboratory (“Argonne”). Argonne, a U.S. 
Department of Energy Office of Science laboratory, is operated under Contract No. DE-AC02-06CH11357. 
Argonne National Laboratory’s work was  funded by the U.S. Department of Energy, Office of High Energy Physics under contract DE-AC02-06CH11357. 
The Askaryan Calorimeter Experiment was supported by the US DOE OHEP under Award Numbers DE-SC0009937, DE-SC0010504, and DE-AC02-76SF0051.
The work of National Central University was funded by the Ministry of Science and Technology, Taiwan.
The work of Korea institutes was supported by the National Research Foundation of Korea (NRF) grants 2020R1A2C3013540, 2021K1A3A1A79097711 and 2018R1A6A1A06024970.


\def\thefootnote{\fnsymbol{footnote}}
\setcounter{footnote}{0}

\bibliographystyle{elsarticle-num}
\def\bibname{\Large\bf References}
\bibliography{references}

\begin{thebibliography}{10}
\expandafter\ifx\csname url\endcsname\relax
  \def\url#1{\texttt{#1}}\fi
\expandafter\ifx\csname urlprefix\endcsname\relax\def\urlprefix{URL }\fi
\expandafter\ifx\csname href\endcsname\relax
  \def\href#1#2{#2} \def\path#1{#1}\fi

\bibitem{Linssen:1425915}
L.~Linssen, A.~Miyamoto, M.~Stanitzki, H.~Weerts, {Physics and Detectors at
  CLIC: CLIC Conceptual Design Report}, CERN Yellow Reports: Monographs, CERN,
  Geneva, 2012, comments: 257 p, published as CERN Yellow Report CERN-2012-003.
\newblock \href {http://dx.doi.org/10.5170/CERN-2012-003}
  {\path{doi:10.5170/CERN-2012-003}}.

\bibitem{Behnke:2013xla}
T.~Behnke, et~al., {The International Linear Collider Technical Design Report -
  Volume 1: Executive Summary}\href {http://arxiv.org/abs/1306.6327}
  {\path{arXiv:1306.6327}}.

\bibitem{FCC:2018evy}
FCC~Collaboration, A.~Abada, et~al., {FCC-ee: The Lepton Collider}: {Future
  Circular Collider Conceptual Design Report Volume 2}, Eur. Phys. J. ST
  228~(2) (2019) 261--623.
\newblock \href {http://dx.doi.org/10.1140/epjst/e2019-900045-4}
  {\path{doi:10.1140/epjst/e2019-900045-4}}.

\bibitem{CEPCStudyGroup:2018ghi}
{CEPC Study Group}, {CEPC Conceptual Design Report: Volume 2 - Physics \&
  Detector}IHEP-CEPC-DR-2018-02, IHEP-EP-2018-01, IHEP-TH-2018-01.
\newblock \href {http://arxiv.org/abs/1811.10545} {\path{arXiv:1811.10545}}.

\bibitem{FCC:2018vvp}
FCC~Collaboration, A.~Abada, et~al., {FCC-hh: The Hadron Collider}: {Future
  Circular Collider Conceptual Design Report Volume 3}, Eur. Phys. J. ST
  228~(4) (2019) 755--1107.
\newblock \href {http://dx.doi.org/10.1140/epjst/e2019-900087-0}
  {\path{doi:10.1140/epjst/e2019-900087-0}}.

\bibitem{Tang:2015qga}
J.~Tang, et~al., {Concept for a Future Super Proton-Proton Collider} (2015).
\newblock \href {http://arxiv.org/abs/1507.03224} {\path{arXiv:1507.03224}}.

\bibitem{Ahmed:2019sim}
Z.~Ahmed, et~al., {New Technologies for Discovery}, in: {CPAD Instrumentation
  Frontier Workshop 2018: New Technologies for Discovery IV (CPAD 2018)
  Providence, RI, United States, December 9-11, 2018}, 2019.
\newblock \href {http://arxiv.org/abs/1908.00194} {\path{arXiv:1908.00194}}.

\bibitem{Cerri:2018rkm}
O.~Cerri, S.~Xie, C.~Pena, M.~Spiropulu, {Identification of Long-lived Charged
  Particles using Time-Of-Flight Systems at the Upgraded LHC detectors}, JHEP
  04 (2019) 037.
\newblock \href {http://arxiv.org/abs/1807.05453} {\path{arXiv:1807.05453}},
  \href {http://dx.doi.org/10.1007/JHEP04(2019)037}
  {\path{doi:10.1007/JHEP04(2019)037}}.

\bibitem{2020chekanov}
S.~V. Chekanov, A.~V. Kotwal, C.~H. Yeh, S.~S. Yu, Physics potential of timing
  layers in future collider detectors, Journal of Instrumentation 15~(09)
  (2020) P09021–P09021.
\newblock \href {http://dx.doi.org/10.1088/1748-0221/15/09/p09021}
  {\path{doi:10.1088/1748-0221/15/09/p09021}}.

\bibitem{2020LLP}
J.~Liu, Z.~Liu, L.-T. Wang, X.-P. Wang, Enhancing sensitivities to long-lived
  particles with high granularity calorimeters at the lhc, Journal of High
  Energy Physics 2020~(11).
\newblock \href {http://dx.doi.org/10.1007/jhep11(2020)066}
  {\path{doi:10.1007/jhep11(2020)066}}.

\bibitem{2014LLP}
Y.~Bai, P.~Schwaller, Scale of dark qcd, Physical Review D 89~(6).
\newblock \href {http://dx.doi.org/10.1103/physrevd.89.063522}
  {\path{doi:10.1103/physrevd.89.063522}}.

\bibitem{2015LLP}
P.~Schwaller, D.~Stolarski, A.~Weiler, Emerging jets, Journal of High Energy
  Physics 2015~(5).
\newblock \href {http://dx.doi.org/10.1007/jhep05(2015)059}
  {\path{doi:10.1007/jhep05(2015)059}}.

\bibitem{THOMSON200925}
M.~Thomson, Particle flow calorimetry and the {PandoraPFA} algorithm, Nuclear
  Instruments and Methods in Physics Research Section A: Accelerators,
  Spectrometers, Detectors and Associated Equipment 611~(1) (2009) 25 -- 40.
\newblock \href {http://dx.doi.org/https://doi.org/10.1016/j.nima.2009.09.009}
  {\path{doi:https://doi.org/10.1016/j.nima.2009.09.009}}.

\bibitem{Sefkow:2015hna}
F.~Sefkow, A.~White, K.~Kawagoe, R.~P\"oschl, J.~Repond, {Experimental Tests of
  Particle Flow Calorimetry}, Rev. Mod. Phys. 88 (2016) 015003.
\newblock \href {http://arxiv.org/abs/1507.05893} {\path{arXiv:1507.05893}},
  \href {http://dx.doi.org/10.1103/RevModPhys.88.015003}
  {\path{doi:10.1103/RevModPhys.88.015003}}.

\bibitem{Tran:2017tgr}
H.~L. Tran, K.~Kr\"uger, F.~Sefkow, S.~Green, J.~Marshall, M.~Thomson,
  F.~Simon, {Software compensation in Particle Flow reconstruction}, Eur. Phys.
  J. C 77~(10) (2017) 698.
\newblock \href {http://arxiv.org/abs/1705.10363} {\path{arXiv:1705.10363}},
  \href {http://dx.doi.org/10.1140/epjc/s10052-017-5298-3}
  {\path{doi:10.1140/epjc/s10052-017-5298-3}}.

\bibitem{CERN-LHCC-2017-023}
\href{https://cds.cern.ch/record/2293646}{{The Phase-2 Upgrade of the CMS
  Endcap Calorimeter}}, Tech. rep., CERN, Geneva (Nov 2017).
\newline\urlprefix\url{https://cds.cern.ch/record/2293646}

\bibitem{Sefkow:2018rhp}
CALICE~Collaboration, F.~Sefkow, F.~Simon, {A highly granular SiPM-on-tile
  calorimeter prototype}, J. Phys. Conf. Ser. 1162~(1) (2019) 012012.
\newblock \href {http://arxiv.org/abs/1808.09281} {\path{arXiv:1808.09281}},
  \href {http://dx.doi.org/10.1088/1742-6596/1162/1/012012}
  {\path{doi:10.1088/1742-6596/1162/1/012012}}.

\bibitem{Graf:2022lwa}
C.~Graf, F.~Simon, {Time-assisted energy reconstruction in a highly-granular
  hadronic calorimeter} (3 2022).
\newblock \href {http://arxiv.org/abs/2203.01317} {\path{arXiv:2203.01317}}.

\bibitem{Akchurin:2021afn}
N.~Akchurin, C.~Cowden, J.~Damgov, A.~Hussain, S.~Kunori, {On the use of neural
  networks for energy reconstruction in high-granularity calorimeters}, JINST
  16~(12) (2021) P12036.
\newblock \href {http://arxiv.org/abs/2107.10207} {\path{arXiv:2107.10207}},
  \href {http://dx.doi.org/10.1088/1748-0221/16/12/P12036}
  {\path{doi:10.1088/1748-0221/16/12/P12036}}.

\bibitem{Baulieu:2015pfa}
G.~Baulieu, et~al., {Construction and commissioning of a technological
  prototype of a high-granularity semi-digital hadronic calorimeter}, JINST
  10~(10) (2015) P10039.
\newblock \href {http://arxiv.org/abs/1506.05316} {\path{arXiv:1506.05316}},
  \href {http://dx.doi.org/10.1088/1748-0221/10/10/P10039}
  {\path{doi:10.1088/1748-0221/10/10/P10039}}.

\bibitem{RevModPhys.90.025002}
S.~Lee, M.~Livan, R.~Wigmans, Dual-readout calorimetry, Rev. Mod. Phys. 90
  (2018) 025002.
\newblock \href {http://dx.doi.org/10.1103/RevModPhys.90.025002}
  {\path{doi:10.1103/RevModPhys.90.025002}}.

\bibitem{2021jettime}
M.~Klimek, The time substructure of jets and boosted object tagging, Journal of
  Physics G: Nuclear and Particle Physics\href
  {http://dx.doi.org/10.1088/1361-6471/ac446a}
  {\path{doi:10.1088/1361-6471/ac446a}}.

\bibitem{Allison2016186}
J.~Allison, et~al., {Recent developments in Geant4}, Nuclear Instruments and
  Methods in Physics Research A 835 (2016) 186.

\bibitem{2017fcc}
S.~V. Chekanov, et~al., Initial performance studies of a general-purpose
  detector for multi-tev physics at a 100 tevppcollider, Journal of
  Instrumentation 12~(06) (2017) P06009–P06009.
\newblock \href {http://dx.doi.org/10.1088/1748-0221/12/06/p06009}
  {\path{doi:10.1088/1748-0221/12/06/p06009}}.

\bibitem{2019granularity}
C.~Yeh, et~al., Studies of granularity of a hadronic calorimeter for
  tens-of-tev jets at a 100 tev pp collider, Journal of Instrumentation 14~(05)
  (2019) P05008–P05008.
\newblock \href {http://dx.doi.org/10.1088/1748-0221/14/05/p05008}
  {\path{doi:10.1088/1748-0221/14/05/p05008}}.

\bibitem{Drasal:20181K}
Z.~Drasal, {Status \& Challenges of Tracker Design for FCC-hh}, PoS Vertex 2017
  (2018) 030.
\newblock \href {http://dx.doi.org/10.22323/1.309.0030}
  {\path{doi:10.22323/1.309.0030}}.

\bibitem{CARTIGLIA201783}
N.~Cartiglia, et~al., Beam test results of a 16ps timing system based on
  ultra-fast silicon detectors, Nuclear Instruments and Methods in Physics
  Research Section A: Accelerators, Spectrometers, Detectors and Associated
  Equipment 850 (2017) 83--88.
\newblock \href {http://dx.doi.org/https://doi.org/10.1016/j.nima.2017.01.021}
  {\path{doi:https://doi.org/10.1016/j.nima.2017.01.021}}.

\bibitem{LGAD21}
E.~L. Gkougkousis, L.~C. Garcia, S.~Grinstein, V.~Coco, Comprehensive
  technology study of radiation hard lgads (2021).
\newblock \href {http://arxiv.org/abs/2111.06731} {\path{arXiv:2111.06731}}.

\bibitem{Benedikt:2018csr}
M.~Benedikt, et~al., \href{https://cds.cern.ch/record/2651300}{{Future Circular
  Collider : Vol. 3 The Hadron Collider (FCC-hh)}}, Tech. Rep.
  CERN-ACC-2018-0058 (2019).
\newline\urlprefix\url{https://cds.cern.ch/record/2651300}

\bibitem{gorham2022picosecond}
P.~W. Gorham, et~al., Picosecond timing-planes for future collider detectors
  (2022).
\newblock \href {http://arxiv.org/abs/2112.00936} {\path{arXiv:2112.00936}}.

\bibitem{PhysRevAccelBeams.21.072901}
P.~W. Gorham, et~al., Picosecond timing of microwave cherenkov impulses from
  high-energy particle showers using dielectric-loaded waveguides, Phys. Rev.
  Accel. Beams 21 (2018) 072901.
\newblock \href {http://dx.doi.org/10.1103/PhysRevAccelBeams.21.072901}
  {\path{doi:10.1103/PhysRevAccelBeams.21.072901}}.

\bibitem{riegler_time_2017}
W.~Riegler, G.~A. Rinella, Time resolution of silicon pixel sensors, J. Inst.
  12~(11) (2017) P11017--P11017.
\newblock \href {http://dx.doi.org/10.1088/1748-0221/12/11/P11017}
  {\path{doi:10.1088/1748-0221/12/11/P11017}}.

\bibitem{Butler:2019rpu}
CMS~Collaboration, J.~N. Butler, T.~Tabarelli~de Fatis, {A MIP Timing Detector
  for the CMS Phase-2 Upgrade}, CERN-LHCC-2019-003, CMS-TDR-020.

\bibitem{Eno2020}
M.~Lucchini, et~al., New perspectives on segmented crystal calorimeters for
  future colliders, Journal of Instrumentation 15~(11) (2020) P11005–P11005.
\newblock \href {http://dx.doi.org/10.1088/1748-0221/15/11/p11005}
  {\path{doi:10.1088/1748-0221/15/11/p11005}}.

\bibitem{Degerli_2020}
Y.~Degerli, et~al., {CACTUS}: a depleted monolithic active timing sensor using
  a {CMOS} radiation hard technology, Journal of Instrumentation 15~(06) (2020)
  P06011--P06011.
\newblock \href {http://dx.doi.org/10.1088/1748-0221/15/06/p06011}
  {\path{doi:10.1088/1748-0221/15/06/p06011}}.

\bibitem{Ronzhin:2014lqa}
A.~Ronzhin, et~al., Development of a new fast shower maximum detector based on
  microchannel plates photomultipliers (mcp-pmt) as an active element, Nucl.
  Instrum. Meth. A 759 (2014) 65--73.
\newblock \href {http://dx.doi.org/10.1016/j.nima.2014.05.039}
  {\path{doi:10.1016/j.nima.2014.05.039}}.

\bibitem{BORTFELDT2020163592}
J.~Bortfeldt, et~al., Timing performance of a micro-channel-plate
  photomultiplier tube, Nuclear Instruments and Methods in Physics Research
  Section A: Accelerators, Spectrometers, Detectors and Associated Equipment
  960 (2020) 163592.
\newblock \href {http://dx.doi.org/https://doi.org/10.1016/j.nima.2020.163592}
  {\path{doi:https://doi.org/10.1016/j.nima.2020.163592}}.

\bibitem{Bortfeldt:2017daz}
THE RD-51 PICOSEC~Collaboration, J.~Bortfeldt, et~al., {PICOSEC: Charged
  particle timing at sub - 25 picosecond precision with a Micromegas based
  detector}, Nucl. Instrum. Meth. A 903 (2018) 317--325.
\newblock \href {http://arxiv.org/abs/1712.05256} {\path{arXiv:1712.05256}},
  \href {http://dx.doi.org/10.1016/j.nima.2018.04.033}
  {\path{doi:10.1016/j.nima.2018.04.033}}.

\bibitem{7581887}
D.~Anderson, et~al., Studies towards a precision timing calorimeter for high
  energy physics collider experiments, in: 2015 IEEE Nuclear Science Symposium
  and Medical Imaging Conference (NSS/MIC), 2015, pp. 1--3.
\newblock \href {http://dx.doi.org/10.1109/NSSMIC.2015.7581887}
  {\path{doi:10.1109/NSSMIC.2015.7581887}}.

\bibitem{CENTISVIGNALI2020162405}
M.~Centis~Vignali, et~al., Deep diffused avalanche photodiodes for charged
  particles timing, Nuclear Instruments and Methods in Physics Research Section
  A: Accelerators, Spectrometers, Detectors and Associated Equipment 958 (2020)
  162405, proceedings of the Vienna Conference on Instrumentation 2019.
\newblock \href {http://dx.doi.org/https://doi.org/10.1016/j.nima.2019.162405}
  {\path{doi:https://doi.org/10.1016/j.nima.2019.162405}}.

\bibitem{Lobanov:2020hhr}
CMS~Collaboration, A.~Lobanov, {Precision timing calorimetry with the CMS
  HGCAL}, JINST 15~(07) (2020) C07003.
\newblock \href {http://arxiv.org/abs/2005.13324} {\path{arXiv:2005.13324}},
  \href {http://dx.doi.org/10.1088/1748-0221/15/07/C07003}
  {\path{doi:10.1088/1748-0221/15/07/C07003}}.

\bibitem{Emberger:2021lsz}
L.~Emberger, {Analysis of Testbeam Data Recorded with the Large CALICE AHCAL
  Technological Prototype}, in: {International Workshop on Future Linear
  Colliders}, 2021.
\newblock \href {http://arxiv.org/abs/2105.08497} {\path{arXiv:2105.08497}}.

\bibitem{app11010111}
Y.~Wang, Y.~Yu, Multigap resistive plate chambers for time of flight
  applications, Applied Sciences 11~(1).
\newblock \href {http://dx.doi.org/10.3390/app11010111}
  {\path{doi:10.3390/app11010111}}.

\bibitem{bib:ultrafast}
C.~Hu, L.~Zhang, R.-Y. Zhu, Ultrafast inorganic crystals with mass production
  capability for future high-rate experiments, contribution to Snowmass 2021.

\bibitem{therrien_energy_2018}
A.~C. Therrien, W.~Lemaire, P.~Lecoq, R.~Fontaine, J.~F. Pratte, Energy
  discrimination for positron emission tomography using the time information of
  the first detected photons, JINST 13~(01) (2018) P01012.
\newblock \href {http://dx.doi.org/10.1088/1748-0221/13/01/P01012}
  {\path{doi:10.1088/1748-0221/13/01/P01012}}.

\bibitem{riegler_time_2021}
W.~Riegler, P.~Windischhofer, Time resolution and efficiency of {{SPADs}} and
  {{SiPMs}} for photons and charged particles, Nucl. Instrum. Meth. A 1003
  (2021) 165265.
\newblock \href {http://dx.doi.org/10.1016/j.nima.2021.165265}
  {\path{doi:10.1016/j.nima.2021.165265}}.

\end{thebibliography}

\end{document}